\documentclass[apj]{emulateapj}
\usepackage{graphicx}
\usepackage{amsmath}
\usepackage{amsfonts}
\usepackage{amssymb}
\usepackage{natbib}
\usepackage{textcomp}
\shortauthors{Chitta et al.}
\begin{document}

\title{Nonlinear Force-Free Field Modeling of the Solar Magnetic Carpet and 
Comparison with \textit{SDO}/HMI and \textit{Sunrise}/IMaX Observations}

\author{L. P. Chitta$^{1,2}$, R. Kariyappa$^1$, A. A. van Ballegooijen$^2$, E. E. DeLuca$^2$, and S.
K. Solanki$^{3,4}$}
\affil{$^1$Indian Institute of Astrophysics, Bangalore 560 034, India}
\affil{$^2$Harvard-Smithsonian Center for Astrophysics, 60 Garden Street MS-58, Cambridge, MA
02138, USA}
\affil{$^3$Max-Planck-Institut f\"ur Sonnensystemforschung, Justus-von-Liebig-Weg 3, 37077
G\"ottingen, Germany}
\affil{$^4$School of Space Research, Kyung Hee University, Yongin, Gyeonggi-Do,446-701, Republic
of Korea}

\begin{abstract}
In the quiet solar photosphere, the mixed polarity fields form a magnetic carpet, which continuously
evolves due to dynamical interaction between the convective motions and magnetic field. This
interplay is a viable source to heat the solar atmosphere. In this work, we used the line-of-sight
(LOS) magnetograms obtained from the Helioseismic and Magnetic Imager (HMI) on the \textit{Solar
Dynamics Observatory} (\textit{SDO}), and the Imaging Magnetograph eXperiment (IMaX) instrument on
the \textit{Sunrise} balloon-borne observatory, as time dependent lower boundary conditions, to
study the evolution of the coronal magnetic field. We use a magneto-frictional relaxation method,
including hyperdiffusion, to produce time series of three-dimensional (3D) nonlinear force-free
fields from a sequence of photospheric LOS magnetograms. Vertical flows are added up to a height of
0.7 Mm in the modeling to simulate the non-force-freeness at the photosphere-chromosphere layers.
Among the derived quantities, we study the spatial and temporal variations of the energy
dissipation rate, and energy flux. Our results show that the energy deposited in the solar
atmosphere is concentrated within 2 Mm of the photosphere and there is not sufficient energy flux at
the base of the corona to cover radiative and conductive losses. Possible reasons and implications
are discussed. Better observational constraints of the magnetic field in the chromosphere are
crucial to understand the role of the magnetic carpet in coronal heating.
\end{abstract}

\keywords{Sun: photosphere --- Sun: magnetic fields --- Sun: atmosphere --- Sun: corona}
\maketitle
\section{INTRODUCTION}\label{sec:intro}
The dynamical evolution of magnetic field in the solar photosphere holds the key to solving the
problem of coronal heating. High resolution observations show that a mixed polarity field, termed as
the magnetic carpet~\citep{1997ApJ...487..424S}, is spread throughout the solar surface. The loops
connecting these elements pierce through the atmosphere before closing down at the photosphere. 
The random motions of these small elements caused by relentless convective motions in the
photosphere, are a favorite candidate to explain the energy balance in the solar atmosphere
\citep[for example, ][]{1998Natur.394..152S,2002ApJ...572L.113G,2002ApJ...576..533P}. In
three-dimensional (3D) magnetohydrodynamic (MHD) models, small-scale footpoint motions drive
dissipative Alfv\'{e}n wave turbulence in coronal loops~\citep{2011ApJ...736....3V}. Also, the
convective motions promote magnetic reconnection and flux cancellation{\textemdash}viable
mechanisms to heat the corona\footnote{The evolution of the magnetic carpet is also studied in the
context of acceleration of slow and fast solar wind
\citep{2010ApJ...720..824C,2013ApJ...767..125C}.} \citep[for
example,][]{1999ApJ...524..483L,2005A&A...439..335G}.

The magnetic carpet typically contains magnetic features with magnetic flux ranging from
$10^{16}-10^{19}$ Mx, a part of which are kilo Gauss flux tubes, also in the internetwork
\citep{2010ApJ...723L.164L}. The carpet is continually recycled with newly emerging flux replacing
the pre-existing flux. Magnetic elements split, merge, and cancel due to granular
action~\citep{2012ApJ...752..149I,2012ApJ...752L..24Y,2013ApJ...774..127L}. Additionally, recent
observations show small-scale swirl events in the chromosphere, possibly due to the rotation of
photospheric elements~\citep{2009A&A...507L...9W}. Also, the magnetic elements display significant
horizontal motions that can reach supersonic speeds~\citep[e.g.][]{2013A&A...549A.116J}.
All these dynamical aspects of the magnetic carpet make the overlying field
non-potential{\textemdash}a source of magnetic energy~\citep[for a review on small-scale
magnetic fields see][]{2009SSRv..144..275D}. There is indirect observational evidence for the
existence of non-potential structures in the quiet Sun~\citep{2013ApJ...778L..17C}.

The vector magnetic field ($\mathbf{B}$) or its line-of-sight component ($B_z$) at the photosphere
is used to infer coronal fields, due to the unavailability of their direct measurements in corona.
Earlier studies of the co-evolution of magnetic carpet and coronal field were mainly through the
potential field (current-free) extrapolations of the photospheric magnetic field 
\citep[for example,][]{2004ApJ...612L..81C,2005ApJ...630..552S}. Recently,
~\citet{2013ApJ...770L..18M} have used nonlinear force-free field extrapolations of the observed
$B_z$ to study the magnetic energy storage and dissipation in the quiet Sun corona. They concluded
that the magnetic free energy stored in the coronal field is sufficient to explain structures like
X-ray bright points (XBP) and other impulsive events at small-scales ~\citep[for reviews on the
force-free magnetic fields see][]{2006SoPh..235..161S,2008SoPh..247..269M,2012LRSP....9....5W}. 
\citet{2013SoPh..283..253W} have used a 22 minute time sequence of very high resolution
vector magnetograms to extrapolate the field into the upper atmosphere under the  potential field
assumption, and argued that the energy release through magnetic-reconnection is not
likely to be the primary contributor to the heating of solar chromosphere and corona in the quiet
Sun. To test the basis of magnetic-reconnection between open and closed flux tubes as a plasma
injection mechanism into the solar wind, \citet{2010ApJ...720..824C} used Monte Carlo simulations of
the magnetic carpet. They also concluded that the slow or fast solar wind is unlikely to be driven
by loop-opening processes through reconnections. The works of \citet{2010ApJ...720..824C} (using
models to study solar wind acceleration), and \citet{2013SoPh..283..253W} (using observations to
study solar atmospheric heating) arrive at similar conclusions, disfavoring a significant role of
the evolution of the magnetic carpet in supplying energy to the corona and solar wind, in contrast
to~\citet{2013ApJ...770L..18M}. 

A possible explanation for the above conflicting results is that the potential field studies
simplify the magnetic topology and do not include dynamic aspects like currents, and other
nonlinear effects. Also, as the methods of analysis are not the same, it may not be so
straightforward to compare the results from those studies. With an ever increasing quality of the
observations showing more intermittent flux filling the solar surface, it is valid to inquire its
contribution to the dynamics of the upper solar atmosphere and advance our knowledge towards a
holistic picture of the magnetic connection from photosphere to corona, particularly in the quiet
Sun.

To rigorously address these issues, a continuous monitoring of the Sun's magnetic field is desired.
This valuable facility is provided by the Helioseismic and Magnetic
Imager~\citep[HMI;][]{2012SoPh..275..207S} on board the \textit{Solar Dynamics
Observatory}~\citep[\textit{SDO};][]{2012SoPh..275....3P}. \textit{SDO}/HMI obtains full-disk
magnetograms of the Sun at $0''.5$ pixel$^{-1}$ with 45 s cadence. To probe the magnetic field at an
even higher spatial resolution, the Imaging Magnetograph
eXperiment~\citep[IMaX;][]{2011SoPh..268...57M} instrument on the \textit{Sunrise} balloon-borne
observatory~\citep{2010ApJ...723L.127S,2011SoPh..268....1B} recorded 33 s cadence observations at
$0''.055$ pixel$^{-1}$. To better understand the role of the magnetic carpet, we use the
\textit{SDO}/HMI and the \textit{Sunrise}/IMaX line-of-sight (LOS) magnetic field observations as
lower boundary conditions in a time-dependent nonlinear force-free modeling of the coronal field,
and report our comparative findings. The rest of the paper is structured as follows. In
Section~\ref{sec:obsvr} we describe the datasets, and model set-up. In Section~\ref{sec:results} the
main results of the work are presented. We conclude in Section~\ref{sec:disc} with a summary, and
implications of the results are discussed. 

\begin{figure*}
\begin{center}
\includegraphics[width=\textwidth]{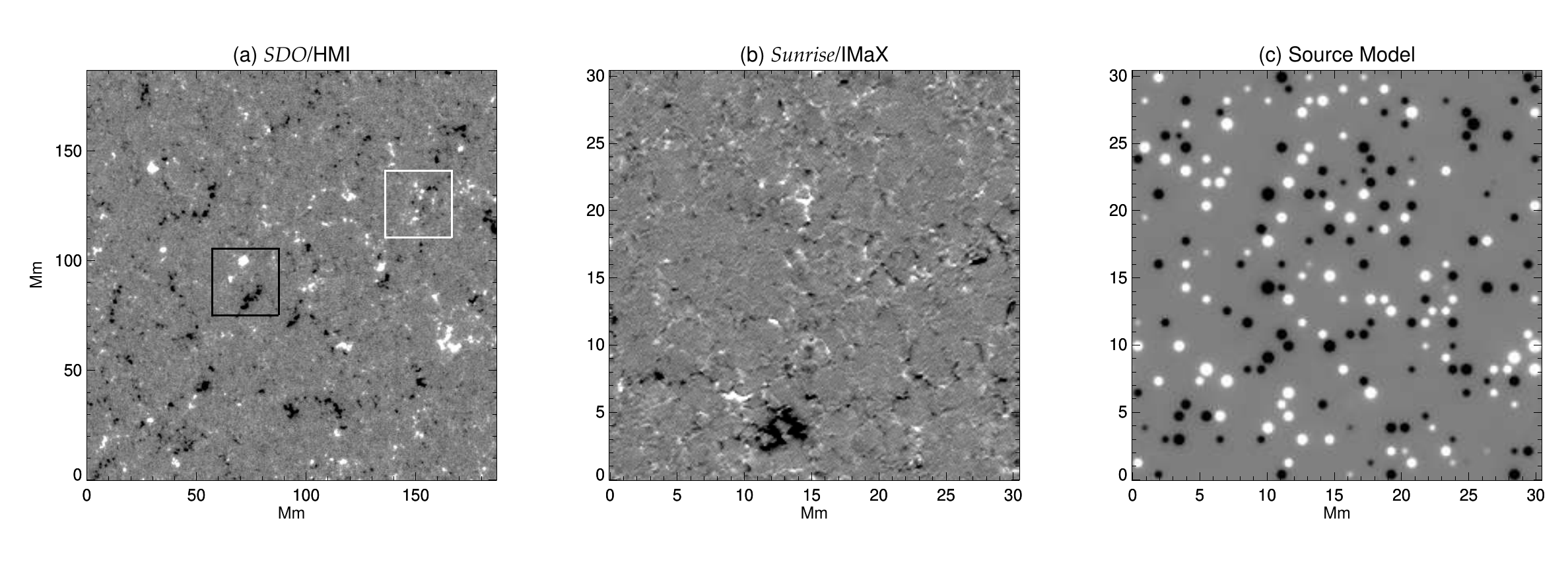}
\caption{Contextual figure showing the photospheric line-of-sight (LOS) magnetograms
taken from the respective time sequences. (a) \textit{SDO}/HMI. (b) \textit{Sunrise}/IMaX. (c)
Model magnetogram from the magnetic sources (see text for details). Note the larger field of view
(FOV) of \textit{SDO}/HMI compared to the \textit{Sunrise}/IMaX, and Source Model.
A white box shown in panel (a) is approximately the size of FOVs in panels (b), and
(c). To facilitate displaying, all the magnetograms are saturated at $\pm50$ G.\label{context}}
\end{center}
\end{figure*}

\section{OBSERVATIONS AND MODEL SET-UP}\label{sec:obsvr}
In this work we used a one day long time sequence of the \textit{SDO}/HMI LOS magnetic field
observations at the disk center and compared the results with the higher resolution observations at
the disk center obtained from the \textit{Sunrise}/IMaX instrument. In Section~\ref{sec:dataset} we
briefly describe the datasets, and Section~\ref{sec:msetup} deals with the set-up of the simulation
for the magnetic field extrapolations.

\subsection{Dataset}\label{sec:dataset}
HMI Data (Set 1): This set consists of a tracked, one day long, time sequence of the LOS
magnetograms observed at the disk center on 2011 January 08, starting at 00:00 UT. With a
field-of-view (FOV) of $187\times187$ Mm$^2$, and a time cadence of 45 s, these observations cover
several magnetic network patches in the quiet Sun. In Figure~\ref{context}(a) we show a snapshot
from the HMI observations. The magnetic flux density is saturated at 50 G. The observations are
found to be closely in flux balance over the entire duration. To retain the weaker field at
the lower boundary, no smoothing is applied to the data. The larger region considered here allows
us to compare various results between sub-regions with varied magnetic configurations. The details
are given in Section~\ref{sec:results}.

\begin{figure*}
\begin{center}
\includegraphics[width=\textwidth]{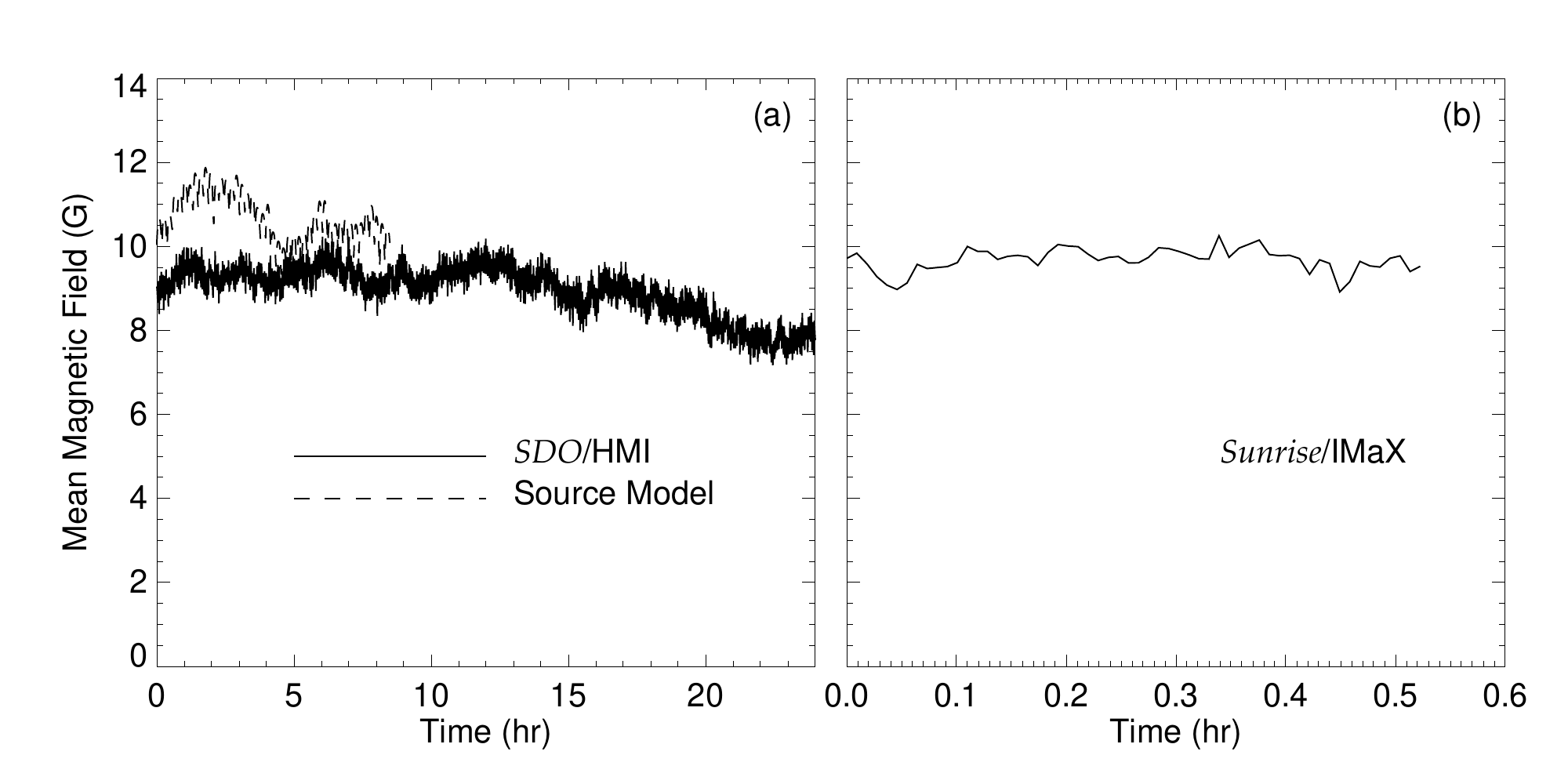}
\caption{Mean magnetic field as a function of time for the three magnetogram
types. (a) \textit{SDO}/HMI (solid line, the mean magnetic field is derived from a region marked
with the white box in Figure~\ref{context}(a)), Source Model (dashed line). (b)
\textit{Sunrise}/IMaX. The $y$-axis scaling for both panels is the same.\label{timeseries}}
\end{center}
\end{figure*}

IMaX Data (Set 2): These level 2 data correspond to imaging spectropolarimetric observations of Fe
{\sc i} at 5250.2 \AA. The observations were recorded at the disk center on 2009 June 09, starting
at 01:30 UT. The spatial resolution of these observations is ten times better than the HMI sequence.
On the downside, IMaX observations span over time period of only $\approx 30$ minutes. From the
original frames, the degraded edges because of the apodization (required by the phase diversity
restoration technique) during the reduction process have been discarded. We extracted a $30\times30$
Mm$^2$ region for further analysis. The LOS magnetograms are derived by taking the ratio of Stokes
$V$ and $I$ with a calibration constant ~\citep[cf. Eq. 17,][]{2011SoPh..268...57M}. The data are
binned to a pixel scale of 119.2 km (Figure~\ref{context}(b)). This effectively reduces the noise in
the data by up to a factor of three (but it also reduces signal due to small-scale mixed polarity
fields). Similar to the HMI data, IMaX data are also found to be closely in flux balance. Due to the
limitations of the time duration of the dataset, the evolution of the coronal magnetic field cannot
be studied for a longer period. For this purpose, we have constructed a series of artificial LOS
magnetograms, named, Source Model. The details of the Source Model are given below.

Source Model (Set 3): The Source Model is made to complement the IMaX observations, but for a longer
duration of time (see Appendix~\ref{source_model} for details). The model is initiated with
$50$ magnetic sources placed randomly on a mesh of hexagonal grids (which has the dimensions of IMaX
FOV considered above). The mesh is periodic in $x$-, and $y$-directions. The sources include both
positive and negative polarity elements such that the net flux is zero. The sources are allowed to
move along the edges of each hexagonal grid which has a length unit of $\approx1$ Mm. All the
sources move with a uniform velocity of 1.5 km s$^{-1}$, which is in the range of typical observed
velocities of the small scale magnetic elements due to the interactions with granules~\citep[for
e.g.][and references therein]{2012ApJ...752...48C}.

During the time evolution, the flux emergence, cancellation, along with splitting and merging of
the elements are the possible ways of interactions among the magnetic elements at the corners of
each grid. The LOS magnetic field reached a quasi-statistical equilibrium almost 15 hours after the
initiation. During this time, the total magnetic flux increased from $10^{19}$ Mx to $10^{20}$ Mx.
Out of the total 48 hours of time evolution, which includes both the rise time and equilibrium
period, a portion of 8.5 hours time sequence is used for further analysis. The total flux
in the Source Model FOV roughly matches with that of the IMaX observations. In
Figure~\ref{context}(c) we show a snapshot of the LOS magnetogram from the Source Model. 

To ensure $\nabla{\cdot}\mathbf{B}=0$, fractional flux imbalance in the observed data (HMI and IMaX)
is corrected by dividing the positive flux with an absolute ratio of integrated positive flux to the
negative flux in the FOV. This method is justified only if the ratio is close to unity to begin
with, in other words, the observations must be closely in flux balance. This is the case in both the
observations we used in this study. In Figure~\ref{timeseries} we plot the mean magnetic field
($\langle|B_z|\rangle$) as a function of time for all the datasets. In the left panel the solid line
is for the HMI set and dashed line is for the Source Model. Right panel shows the IMaX mean magnetic
field. We emphasize that the IMaX LOS magnetic field is derived from a simple ratio method, which
underestimates the flux density in stronger elements~\citep{2011SoPh..268...57M}. 

\begin{deluxetable*}{lccccc} 
\tablecolumns{5} 
\tablewidth{0pc} 
\tablecaption{Values of $L$, $T$, $\eta_4$ and $\nu^{-1}$ used in the magnetic modeling
\label{tab:tab1}} 
\tablehead{ 
\colhead{} & $L$ (Mm) & $T$ (s)  &  $\eta_4$ (km$^4$ s$^{-1}$) & $\nu^{-1}$ (km$^2$ s$^{-1}$) 
}
\startdata 
HMI          & 3.12 & 900 & $1.05\times10^7$  & $1\times10^3$   \\
IMaX         & 1.02 & 165 & $6.48\times10^6$  & $0.6\times10^3$ \\
Source Model & 1.02 & 680 & $1.57\times10^6$  & $0.1\times10^3$ \\
\enddata
\end{deluxetable*} 
   
\subsection{Simulation Set-up}\label{sec:msetup}
In this section we describe the simulation set-up and the equations we solve to derive the 3D
magnetic field above the photosphere. 

For Set 1, the computational box is a $512\times512\times250$ cell volume covering
$187\times187\times91.4$ Mm$^3$ in physical space. For Sets 2 and 3, the computational domain is
much smaller covering only $30.5\times30.5\times18$ Mm$^3$ with $256\times256\times150$ cells.
The simulation box is periodic in $x$- and $y$-directions and closed at the top. LOS magnetograms
of Sets 1, 2, and 3 are the respective boundary conditions in the $xy$-plane at $z=0$. Initially,
at time $t=0$, a potential field is assumed to fill the box, which is then evolved in time into
nonlinear force-free states with the evolving boundary conditions.

The rate of change of vector potential $\mathbf{A}$ is related to the magnetic field $\mathbf{B}$
through the induction equation
\begin{equation}
 \frac{\partial\mathbf{A}}{\partial t}=\mathbf{v}\times\mathbf{B}+\epsilon \label{eq:ind},
\end{equation}
where $\epsilon$ is the hyperdiffusion, defined as 
\begin{equation}
 \epsilon=\frac{\mathbf{B}}{B^2}\nabla\mathbf{\cdot}(\eta_4B^2\nabla\alpha).\label{eq:hd}
\end{equation}
$\alpha={\mathbf{j{\cdot}B}}/{B^2} $ is the force-free parameter, and $\eta_4$ is the
hyperdiffusivity. Some form of magnetic diffusion is needed in any numerical MHD simulation
to suppress the development of numerical artifacts on small spatial scales. Here we use
hyperdiffusion (instead of ordinary resistivity) to minimize the direct effect of the diffusion on
larger scales. \citet{2008ApJ...682..644V} presented a theory of coronal heating which draws energy
with hyperdiffusion from the dissipation of nonpotential magnetic field. It conserves the mean
magnetic helicity and smooths the gradients in
$\alpha$~\citep{1986JPlPh..35..133B,1986PhRvL..57..206B}. It has been proposed that tearing modes
can promote turbulence in 3D sheared magnetic field, which causes
hyperdiffusion~\citep{1988ApJ...326..412S}.

Equation~\ref{eq:ind} is evolved using a magneto-frictional relaxation
technique~\citep{1986ApJ...309..383Y}, which assumes that the plasma velocity ($\mathbf{v}$, in
this case, the magneto-frictional velocity) is proportional to the Lorentz force
($\mathbf{j}\times\mathbf{B}$), given by
\begin{equation}
 \mathbf{v}=\frac{1}{\nu}\frac{\mathbf{j}\times\mathbf{B}}{B^2}, \label{eq:mfr1}
\end{equation}
where $\nu$ is the frictional coefficient. The numerical values of $\eta_4$ and $\nu^{-1}$
(tabulated in Table~\ref{tab:tab1}) are determined in part by the requirement
of numerical stability of the code, and by the length ($L$) and time ($T$) units of the
simulations. \citet{2011ApJ...729...97M} and~\citet{2012ApJ...757..147C} used
magneto-frictional relaxation technique to study the time evolution of active regions on much
larger scales.

It is known that the magnetic field in the photosphere is non-force-free due to the high plasma
$\beta$.~\citet{1995ApJ...439..474M} calculated the Lorentz force of an active region as a function
of height. They concluded that the field becomes force-free higher in the atmosphere, above
approximately 400 km. The mixed polarity fields in the magnetic carpet are weaker and may be
dominated by gas pressure even in the chromosphere as suggested by the simulations
of~\citet{2007ApJ...665.1469A}, which, however, do not include a realistic treatment of radiative
transfer for the photospheric and chromospheric layers. To mimic the additional forces on
the magnetic field in the photosphere and low chromosphere, we use the method first described by
\citet{2008SoPh..247..269M} where a second term is added to the magnetofrictional velocity
\begin{equation}
 \mathbf{v}=(\frac{1}{\nu}\mathbf{j}-v_1\hat{z}\times\mathbf{B})\times\mathbf{B}/B^2.\label{eq:mfr2}
\end{equation}
The additional term is present only at low heights (up to 0.7 Mm) and is the projection of a
constant upward velocity ($v_1 \hat{z}$) onto the plane perpendicular to the local magnetic field.
The quantity $v_1$ is constant over horizontal planes and independent of the magnetic field. This
new force marginally prevents the field from splaying out at the lower boundary. In the present
paper we use $v_1=1.5$ km s$^{-1}$ for IMaX/Source Model, and $v_1=3.4$ km s$^{-1}$ for HMI dataset.
This has only a minor effect on the flux concentrations in our model. Similar to $\eta_4$ and
$\nu^{-1}$, the choice of $v_1$ is defined by the $L$ and $T$ units of the respective models.

\begin{figure*}%
\begin{minipage}{\textwidth}%
\begin{center}
\includegraphics[width=\textwidth]{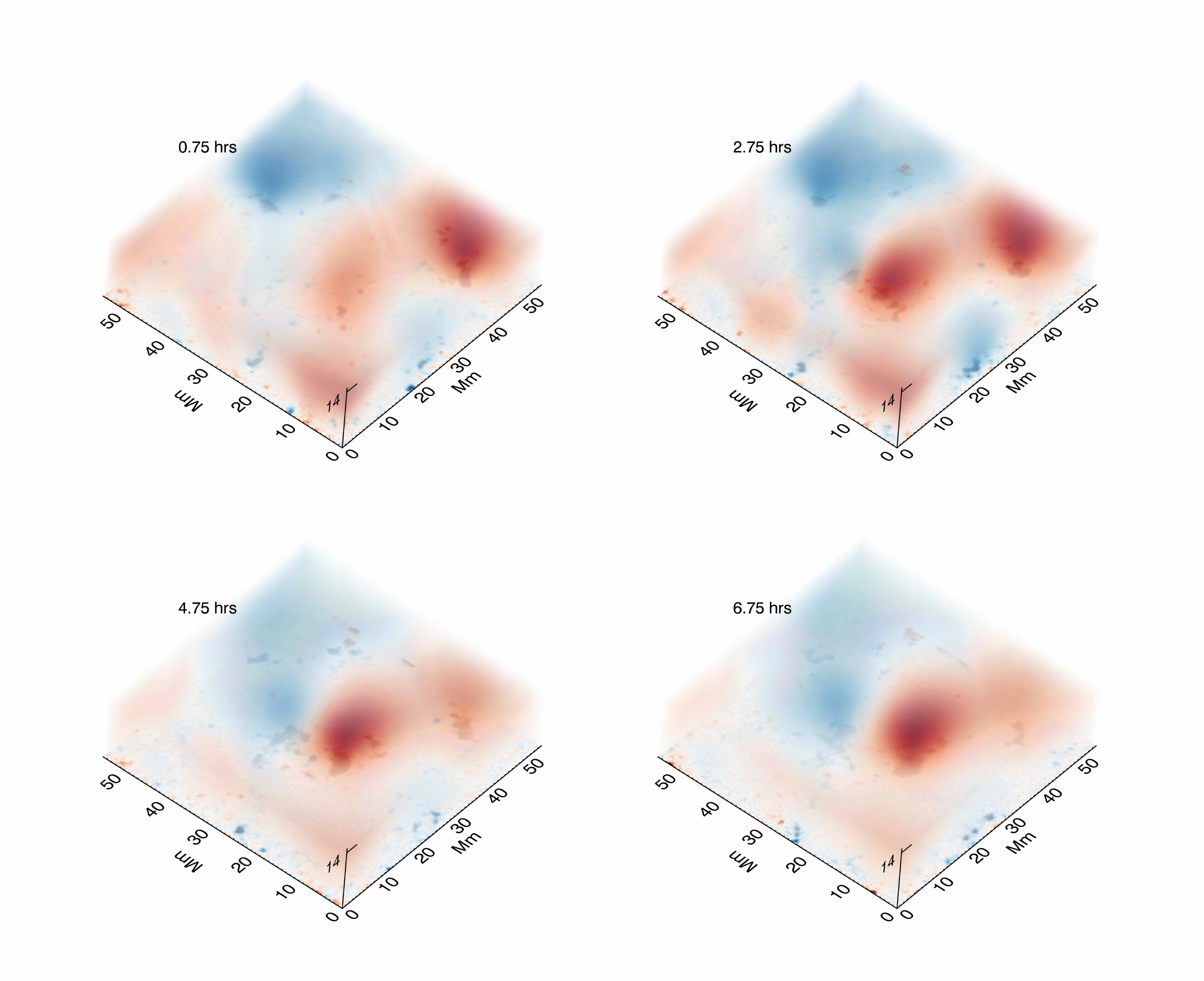}\\
\includegraphics[width=\textwidth]{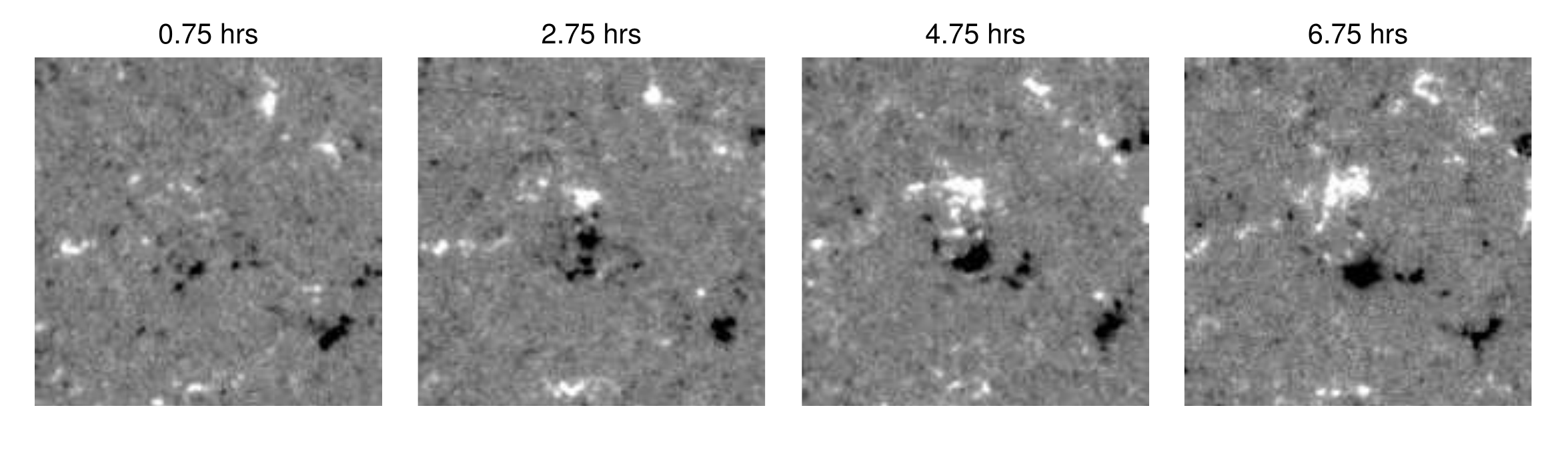}
\end{center}
\end{minipage}
\caption{Top segment: Volume rendering of the extrapolated $B_z$ using \textit{SDO}/HMI observations
as lower boundary conditions. Shown here is the time evolution of $B_z$ of an ephemeral region, with
consecutive frames lying two hours apart, in a box of $\approx 55\times 55\times 14$ Mm$^3$ volume,
with its bottom surface centered at (70 Mm, 90 Mm) of Figure~\ref{context}(a). The red (blue)
colored areas are the regions of negative (positive) polarity. Bottom segment: The distribution
of $B_z$ (saturated at $\pm50$ G) in the photosphere corresponding to the respective panels in the
top segment is shown.}
\label{render} 
\end{figure*}

\section{RESULTS}\label{sec:results}
In the top segment of Figure~\ref{render} we present volume rendering of $B_z$ in a sub-volume from
HMI simulations, at four instances, separated by two hours each. At the lower boundary of this
sub-volume an emerging bipole is seen, with stronger negative polarity (centered on the black
square, Figure~\ref{context}(a)). The red (blue) coloration covers regions of negative (positive)
polarity. As the bipole emerges into the atmosphere, pre-existing coronal field responds to it, and
eventually becomes bipolar in time. Note that this is a 3D rendering. To \textit{see through} the
cube, we used increasing transparency of layers with height (so that we can see till the bottom
through the layers above). Also, to account for the decrease in field strength with height, we
scaled each layer (at a given height) in respective panels separately. In the bottom segment
we show $B_z$ at photospheric level corresponding to the respective panels in the top segment. The
purpose of this figure is to show how the coronal field responds to an emerging bipole in the
photosphere.

\begin{figure}
 \begin{center}
  \includegraphics[width=0.475\textwidth]{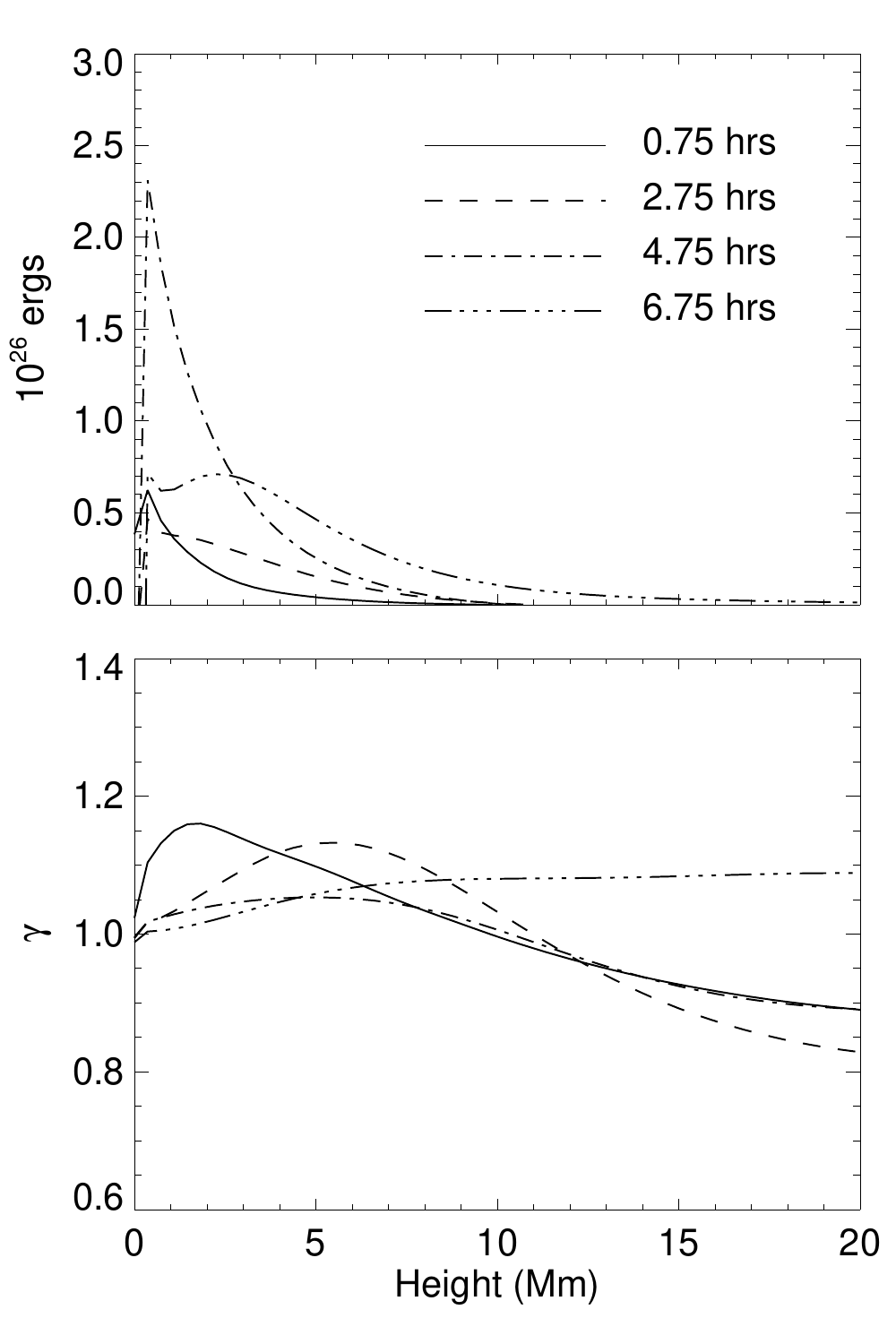}
  \caption{Top panel: Free energy of an emerging bipole located within the black square in
Figure~\ref{context}(a). Each of the four curves plotted in a different line style is for a
different point in time, corresponding to the times of the snapshots plotted in Figure~\ref{render}.
The curves are labelled in the upper panel. At all times $E_{\text{free}}(z)$ falls rapidly with
$z$, peaking below 5 Mm. Bottom panel: The ratio of non-potential magnetic energy to the potential
magnetic energy ($\gamma$) is plotted as a function of height (for the same time steps as shown in
the top panel). The peaks of $\gamma(z)$ shift towards higher $z$ with time, indicating the
interaction of newly emerging field with the pre-existing coronal field. \label{free_ratio}}
 \end{center}
\end{figure}

The magnetic free energy ($E_{\text{free}}$), which is an excess of non-potential magnetic
energy (in this case, the nonlinear force-free energy) over the potential magnetic energy is
calculated. Our aim is to emphasize the height dependence of the quantities relevant to the energy
budget of the solar corona. To this end, we calculate $E_{\text{free}}$ as a function of height,
given by
\begin{equation}
 E_{\text{free}}(z) = \frac{1}{8\pi}\Delta z \int \int
[B^2_{\text{np}}(x,y,z)-B^2_{\text{p}}(x,y,z)]dxdy, 
\end{equation}
where $\Delta z$ is the pixel length. $B^2_{\text{np}}/8\pi$ and $B^2_{\text{p}}/8\pi$ are the
non-potential, and potential magnetic energy densities, respectively. In the top panel of
Figure~\ref{free_ratio} we plot the HMI $E_{\text{free}}(z)$ ($\Delta z=0.36$ Mm). Integration is
over the surface enclosed by the black square in Figure~\ref{context}(a). The curves represent the
evolution of an emerging bipole at four times (shown in Figure~\ref{render}). It is observed that
most of $E_{\text{free}}(z)$ (5\textendash 10$\times10^{25}$ ergs), is available at heights below 5
Mm. These results are consistent with the free energy values reported
in~\citet{2013ApJ...770L..18M}. 

To estimate the non-potentiality of the magnetic energy, we calculate a ratio $\gamma(z)$, defined
as
\begin{equation}
 \gamma(z) = \frac{\int \int B^2_{\text{np}}(x,y,z)dxdy}{\int \int B^2_{\text{p}}(x,y,z)dxdy}.
\end{equation}
It is observed that the magnetic energy is close to energy of the potential magnetic field, with an
excess $<20$\%. In the bottom panel of Figure~\ref{free_ratio} we plot $\gamma(z)$ for the same time
steps as shown in the top panel. There is an apparent rise in the non-potentiality with time,
indicating the interaction of a newly emerging flux with the pre-existing coronal field. However, by
combining the values of $\gamma(z)$ and $E_{\text{free}}(z)$, it can be seen that the coronal field
over the HMI FOV is potential. We note that in the bottom panel, $\gamma(z)<1$ for some
points in time above a height of $10$ Mm, which means a \textit{negative} free energy. This is due
to the fact that the surface integration is performed only over a sub-region within the full FOV.
The total volume integrated $E_{\text{free}}$, however, remains positive.

\begin{figure*}
\begin{center}
\includegraphics[width=\textwidth]{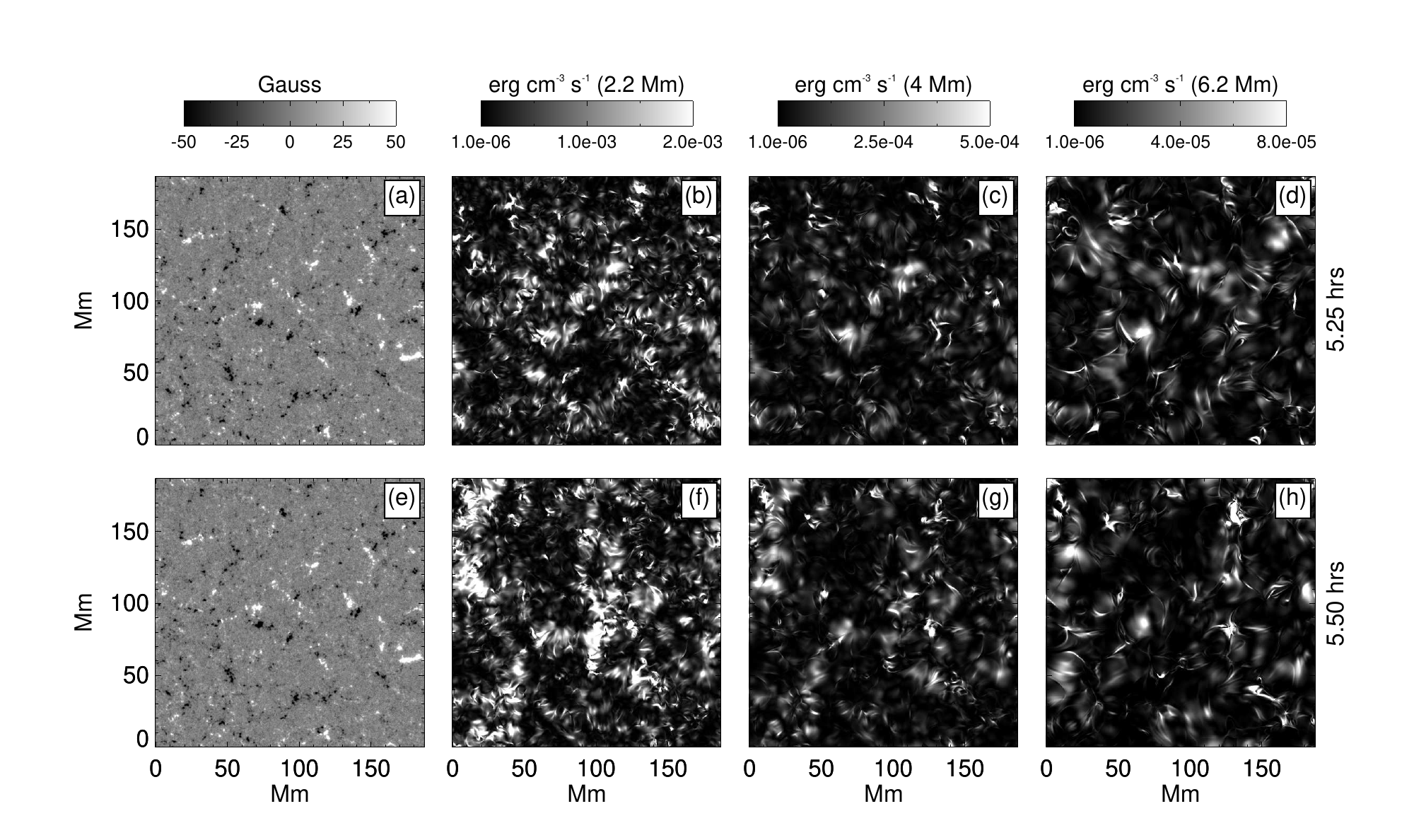}
\caption{HMI Magnetic field configuration at the lower boundary ((a) and (e) saturated at $\pm50$
G), and the energy dissipation ($Q$) at three layers ((b) and (f) 2.2 Mm; (c) and (g) 4 Mm; (d) and
(h) 6.2 Mm) are shown. Note the change of grey scale ranges in the energy dissipation maps. The top
and the bottom maps are separated by 15 minutes in time and are rendered according to the same
greyscales (for a given height). Though the magnetic field configuration has little apparent change,
the locations of the energy dissipation have changed significantly, suggesting a highly
\textit{non-localized} and intermittent nature of the time evolution of magneto-frictional
relaxation method.\label{sdo_disp}}
\end{center}
\end{figure*}

Magnetic free energy is necessary but not sufficient to completely quantify the energy supply to
solar corona. It is important to calculate the energy dissipation, and energy flux, which can then
be directly compared with the observed radiative and conductive losses. Earlier studies by
\citet{1977ARA&A..15..363W}, \citet{1988ApJ...325..442W}, \citet{1991ApJ...382..667H} placed
observational constraints on the energy flux through the coronal base, $10^5$\textendash $10^6$ erg
cm$^{-2}$ s$^{-1}$ for a quiet Sun region. With magneto-frictional relaxation including
hyperdiffusion, the energy dissipation per unit volume ($Q$) is calculated as 
\begin{equation}
Q=\frac{B^2}{4\pi}(\nu|\mathbf{v}|^2+\eta_4|\nabla\alpha|^2), \label{eq:disp}
\end{equation}
\citep[see,][]{1986ApJ...309..383Y,2008ApJ...682..644V}. The first and second parts on the right
hand side of the Equation~\ref{eq:disp} are due to magneto-friction, and hyperdiffusion,
respectively. To understand the evolution of $Q$ with respect to height, at the time scale of
minutes, we consider snapshots of $Q$ from HMI simulations, separated by 15 minutes. The results are
shown in Figure~\ref{sdo_disp}. The top row shows $B_z$ (lower boundary, (a)), and $Q$ at
three heights ((b): 2.2 Mm, (c): 4 Mm, and (d): 6.2 Mm) at 5.25 hrs into the simulation. Bottom row
is same as the top row, but 15 minutes after panels (a)-(d). With very little change in the $B_z$
over the duration, apparently, there is a drastic morphological change in $Q$ (panels (b) and (f)).
This is very interesting to note, suggesting that the energy dissipation process due to
magneto-friction and hyperdiffusion is \textit{non-localized} in time (i.e., the locations of energy
dissipation rapidly change with time). This in turn means that the time evolution of spatially
averaged $Q$ at any two locations separated by some distance (having a different underlying magnetic
morphology) would be similar. In other words, the average dissipation of magnetic energy is very
uniform in our simulations. However, the magnitude of $Q$ has a strong height dependence (as seen in
the greyscales of the right three panels). This was also indicated by~\citet{2013ApJ...770L..18M}.
These large-scale changes in the atmosphere are caused by small-scale random footpoint motions of
magnetic elements at the photosphere.

\begin{figure*}
\begin{center}
\includegraphics[width=\textwidth]{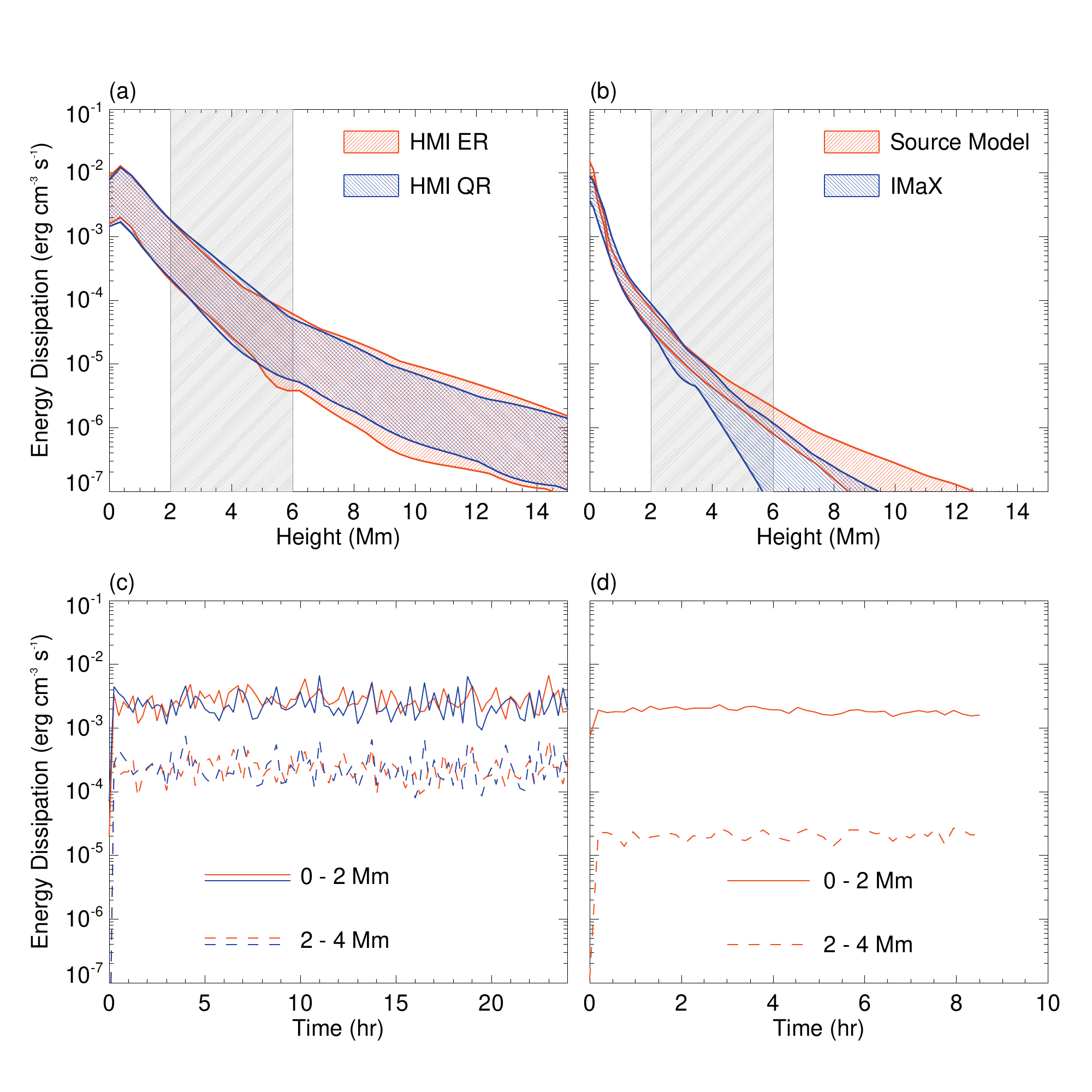}
\caption{Horizontal mean of energy dissipation plotted as a function of height (top panels), and
time (bottom panels). (a) Mean energy dissipation over an ER (red), and a QR (blue) observed by HMI.
(b) Energy dissipation in the IMaX dataset (blue) and Source Model (red) averaged over the entire
FOV. As an illustration, a grey striped box is drawn at 2\textendash 6 Mm range to compare the
energy dissipations between panels (a), and (b) at those heights. (c) Mean energy dissipation
plotted vs. time, averaged over the height ranges 0\textendash 2 Mm (solid) and 2\textendash 4 Mm
(dashed), respectively. (d) Same as (c), but for Source Model. \label{dissipat}}
\end{center}
\end{figure*}

To further quantify the above statements, Figure~\ref{dissipat} plots the horizontally averaged
dissipation rate as a function of height and time. The top left panel is $Q$ plotted as a function
of height from the HMI simulations. The two shaded bands are for two sub-regions of the HMI FOV
(red: Figure~\ref{context}(a) black square; blue Figure~\ref{context}(a) white square). The black
square covers an ephemeral region (ER), and the white square covers a \textit{less active} quiet
region (QR). At a given height, the width of each shaded band represent the range of minimum to
maximum energy dissipated at that height. A vertical shaded rectangle is drawn to indicate the base
of the transition region and corona. $Q$ monotonically decreases with height in a similar way for
regions with completely different underlying $B_z$. In the top right panel, results from the Source
Model (red), and IMaX (blue) are shown for comparison with those of HMI. It is noted that the fall
of $Q_{\text{Source Model/IMaX}}$ is steeper than $Q_{\text{HMI}}$.

\begin{figure*}
\begin{center}
\includegraphics[width=\textwidth]{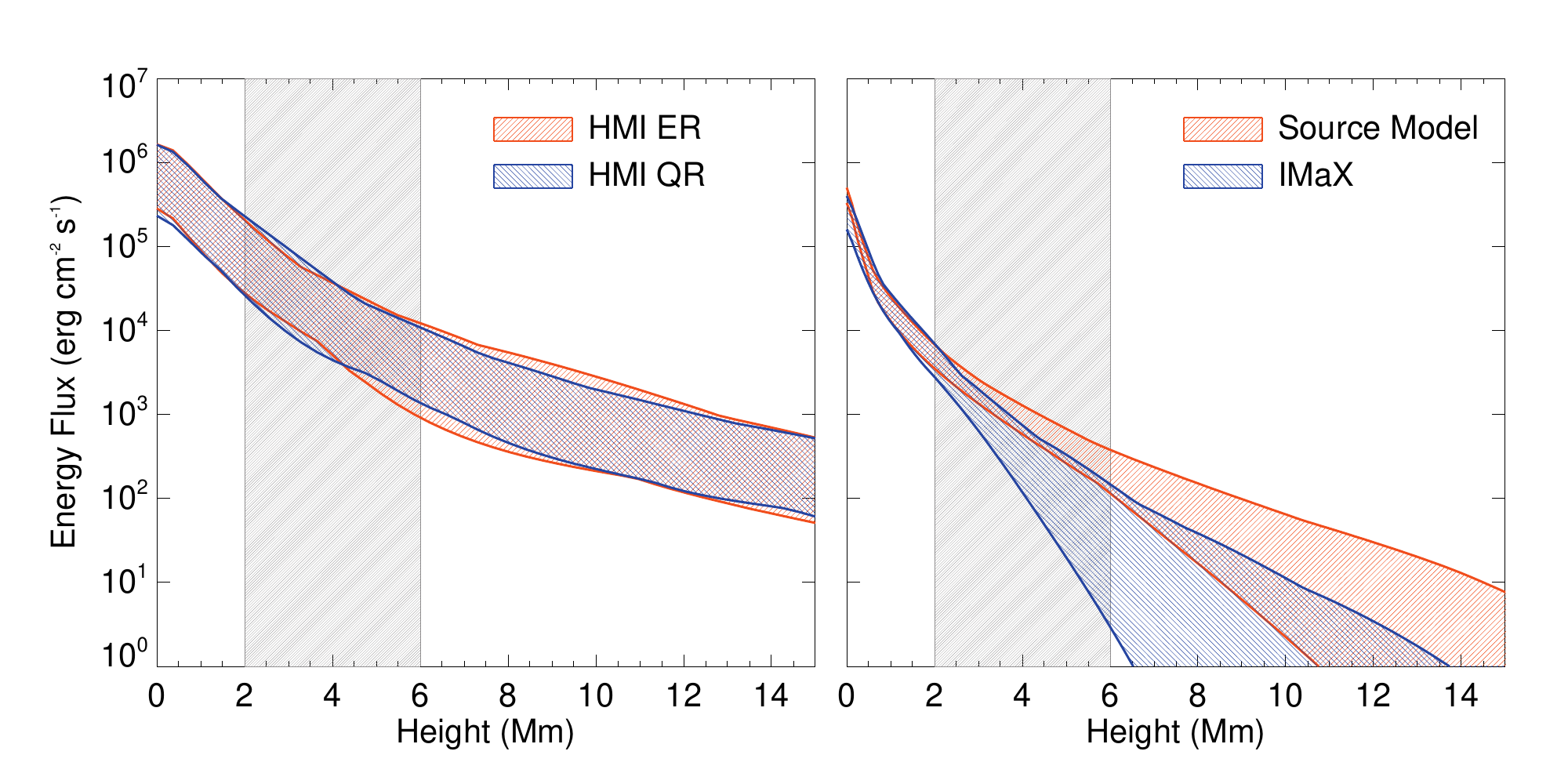}
\caption{Horizontal mean of the energy flux plotted as function of height. The color schemes and
legends in the left and right panels are the same as in Figures~\ref{dissipat}(a) and (b),
respectively.
\label{eng_flux}}
\end{center}
\end{figure*}

Further, we plot the averaged $Q$ in two height ranges (0\textendash 2; 2\textendash 4 Mm) and study
its time evolution. In Figure~\ref{dissipat}(c) we plot these results for the HMI simulations (red:
ER; blue: QR; solid: 0\textendash 2 Mm; dashed: 2\textendash 4 Mm). These plots show that on
average, the energy dissipation is uniform over the entire FOV and also fairly constant throughout
the time evolution. Another fact to notice is that $Q$ reaches its statistically mean value very
early in the simulations. Recalling that the initial condition of the simulation is a potential
field, the build-up of magnetic shear of coronal field and in turn the energy dissipation through
hyperdiffusion, is contributing only a fraction of total $Q$. Magneto-friction acts as a dominant
energy source in our simulations. Figure~\ref{dissipat}(d) shows the evolution of $Q_{\text{Source
Model}}$. 

As a next step, we calculate the energy flux ($F$) through a given horizontal surface. In a
quasi-stationary state, the energy flux $F(z)$ through height $z$ is equal to the integral
of $Q(z)$ over all heights above $z$, given by
\begin{equation}
 F(z) = \int_z^{z_{max}} Q(z) dz,
\end{equation}
where $z_{max}$ is the top of the simulation domain. The horizontal average of the calculated $F(z)$
is plotted in Figure~\ref{eng_flux}. Similar to Figure~\ref{dissipat}(a), in the left panel we show
$\langle F(z)\rangle$ for two sub-regions within HMI FOV. $\langle F(z)\rangle$ of IMaX/Source Model
are plotted in the right panel. All sets have a flux of $\approx 10^6$ erg cm$^{-2}$ s$^{-1}$ at
the photosphere. Almost all of this flux is dissipated at heights below 2 Mm, i.e. in the
chromosphere (for which, however, our model may not be valid, since the field below
1000\textendash 1500 km is not force-free). Over a quiet Sun, the coronal base is typically above 3
Mm~\citep[cf. Figure 3,][]{1977ARA&A..15..363W}. At this layer, $\langle F_{\text{HMI}}(z)\rangle
\approx 5\times10^4$ erg cm$^{-2}$ s$^{-1}$, and $\langle F_{\text{IMaX}}(z)\rangle \approx
2\times10^3$ erg cm$^{-2}$ s$^{-1}$. These values are lower than the required flux in the quiet Sun
corona. 

To summarize Figures~\ref{dissipat} and~\ref{eng_flux}, three sets of magnetograms used in this
work show a similar trend. Close to the lower boundary both $Q(z)$, and $F(z)$, respectively, are
similar for all the cases. $F(z)$ is $>10^5$ erg cm$^{-2}$ s$^{-1}$, and appears to be comparable
with the coronal energy budget of the quiet Sun. This condition is no more satisfied for
$z>$2\textendash 3 Mm. On average, there is an order of magnitude or more deficit in the required
energy flux to support the observed coronal energy losses. The deficit in $F(z)$ is very noticeable
in the higher resolution but smaller FOV IMaX/Source Model cases.

\begin{figure}
\begin{center}
\includegraphics[width=0.475\textwidth]{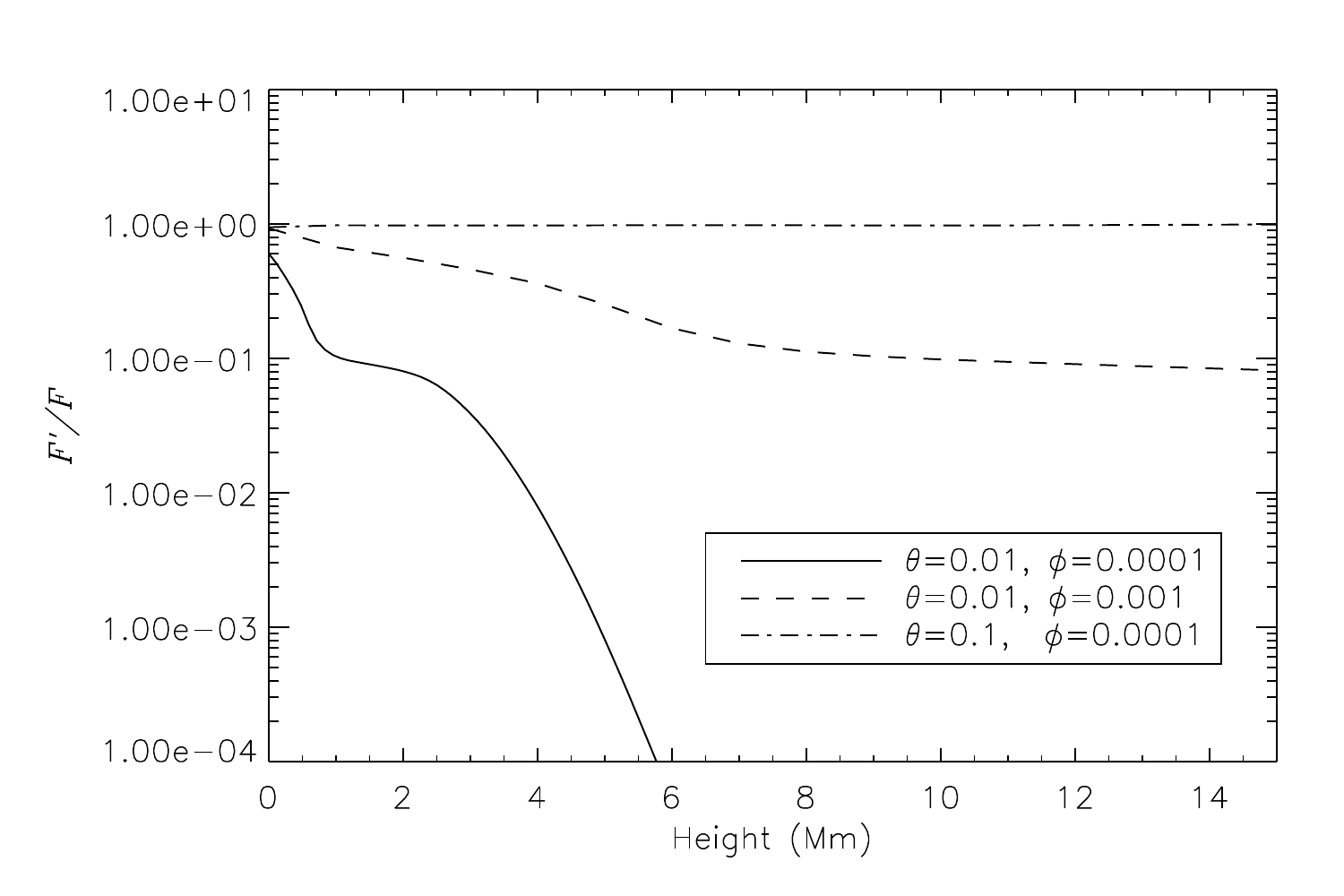}
\caption{The dependence of energy flux on the dimensionless parameters ($\theta$, $\phi$) is
shown for the case of IMaX dataset. The ratio $F^\prime_{\theta,\phi} (z) / F(z)$ is
plotted as a function of height for three sets of parameters as marked in the plot legend (see
text for details).\label{eta_nu}}
\end{center}
\end{figure}

Next, choosing IMaX set as an example we studied the dependence and sensitivity of the
calculated energy dissipation and flux on the simulation parameters ($\nu^{-1}$, $\eta_4$). As
described in the previous section, the numerical values of these parameters are constrained by the
requirements that the code be numerically stable and that numerical artifacts are suppressed. The
frictional coefficient and hyperdiffusivity can be written as $\nu^{-1} = \theta L^2/T$ and $\eta_4
= \phi L^4/T$, where $\theta$ and $\phi$ are dimensionless parameters, and $L$ and $T$ are the
length and time units for each simulation. For the case of IMaX (and also for the Source Model),
$\theta  = 0.1$ and $\phi = 0.001$; the corresponding values of $\nu^{-1}$ and $\eta_4$ are listed
in Table~\ref{tab:tab1}. Numerical stability dictates that $\theta$ and $\phi$ cannot be arbitrarily
large, and to suppress numerical artifacts $\phi$ cannot be arbitrarily small. To test how lower
values of these parameters affect the final results, we considered three sub-cases: (a) $\theta =
0.01$, $\phi = 0.0001$; (b) $\theta = 0.01$, $\phi = 0.001$; (c) $\theta = 0.1$, $\phi = 0.0001$.
Further, the energy flux with the respective new set of parameters is calculated, and we denote it
as $F^\prime_{\theta,\phi} (z)$. The ratio $F^\prime_{\theta,\phi} (z) / F(z)$ is plotted in
Figure~\ref{eta_nu} (solid line: case (a); dashed line: case (b); dash-dotted line: case(c)). 
Here, $F(z)$ is calculated with $\theta  = 0.1$ and $\phi = 0.001$ (see the blue shaded band
in the right panel of Figure~\ref{eng_flux}). These results indicate that the energy dissipation and
flux are sensitive to the frictional coefficient, which can also be argued on the basis that the
magneto-frictional energy is dominant in the simulations.

\begin{figure*}%
\begin{minipage}{\textwidth}%
\begin{center}
\includegraphics[width=0.49\textwidth]{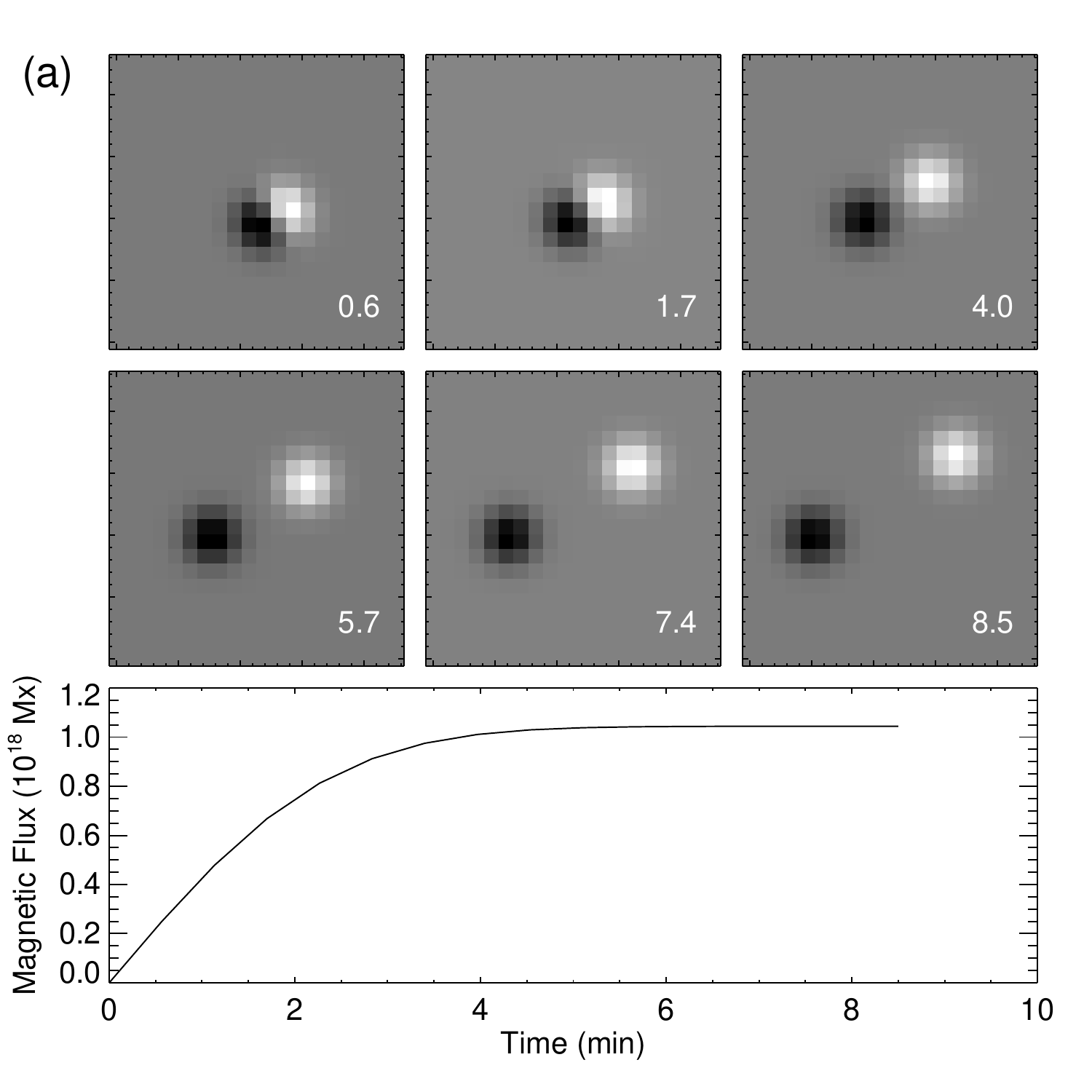}
\includegraphics[width=0.49\textwidth]{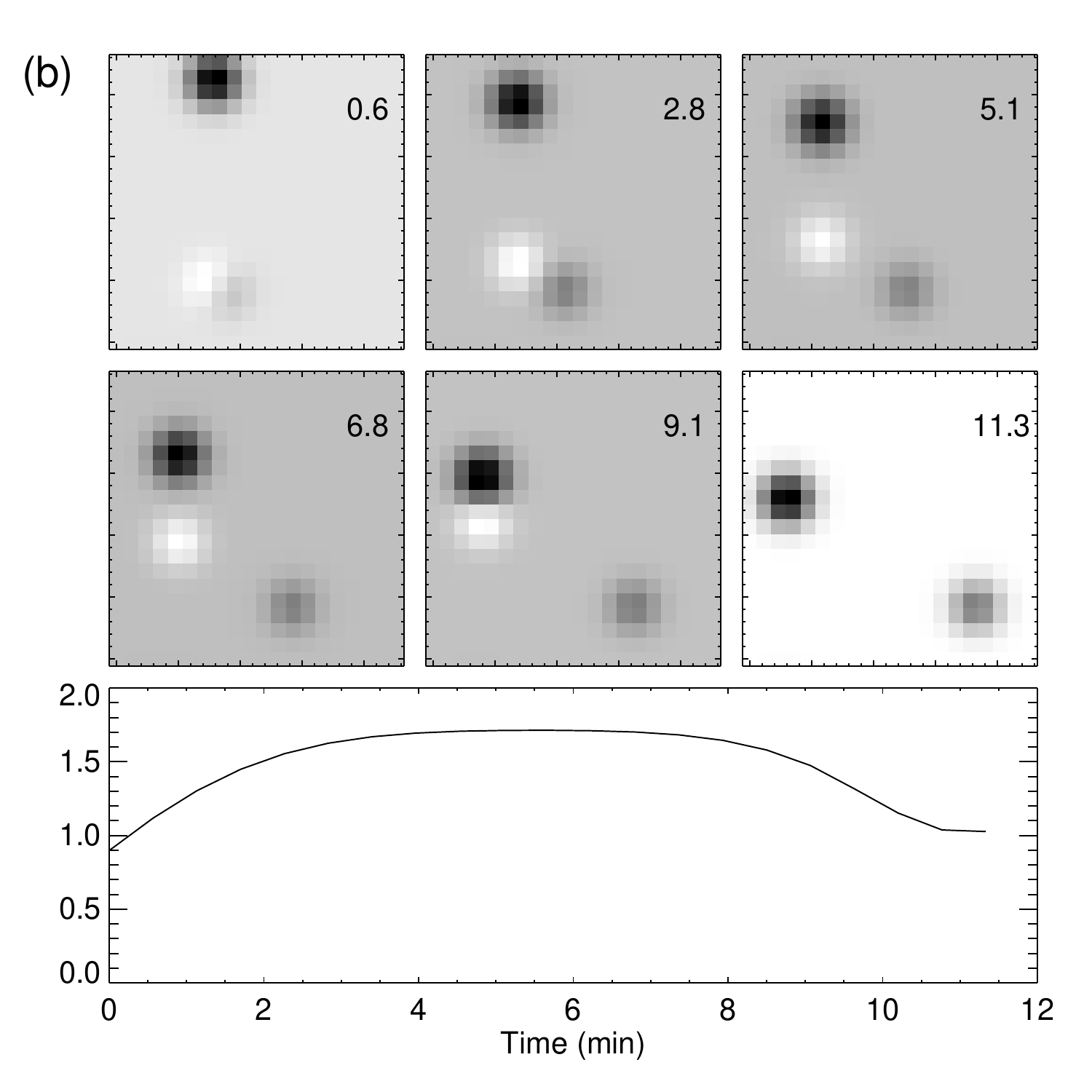}\\
\includegraphics[width=0.49\textwidth]{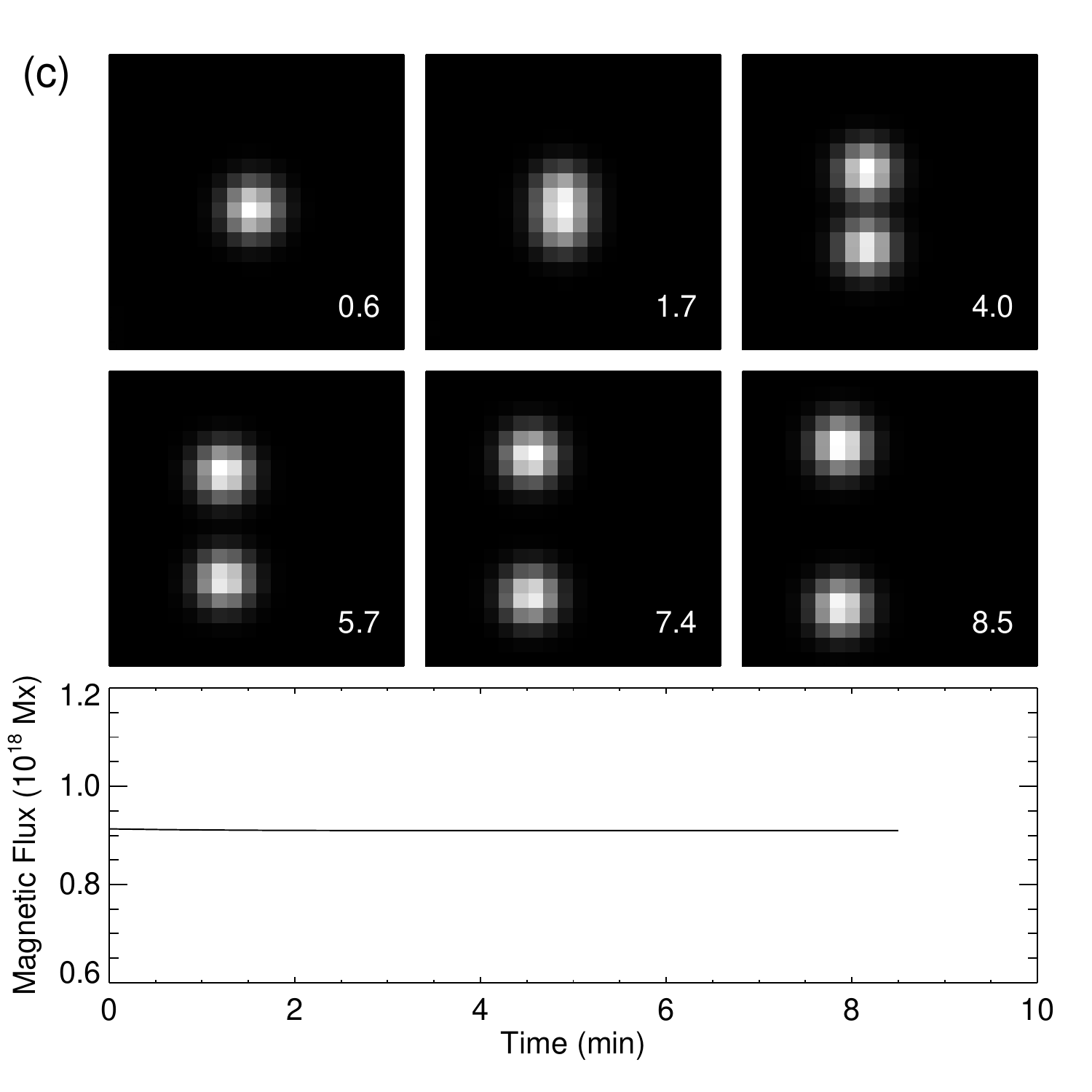}
\includegraphics[width=0.49\textwidth]{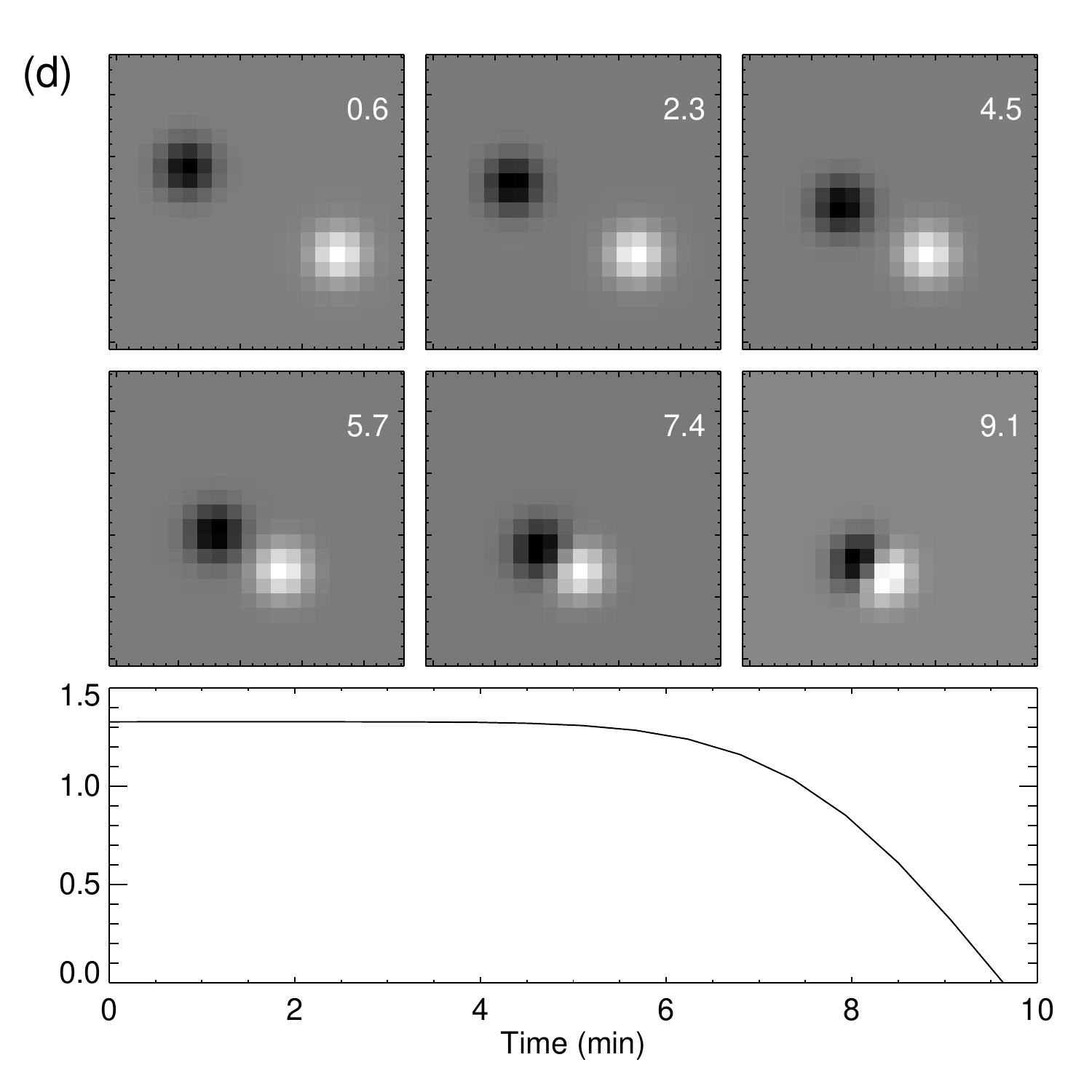}
\end{center}
\end{minipage}%
\caption{Various interactions among magnetic elements in the Source Model are illustrated here. (a)
Flux emergence. (b) Flux emergence followed by a partial flux cancellation. (c) Flux splitting. (d)
Total flux cancellation. Each panel is comprised of two segments. The top segment is a time
sequence of magnetic maps describing the interaction. Each map in the respective top segment
covers an area of $\approx2.4\times 2.4$ Mm$^2$. The numbers in the maps indicate time, in
minutes, elapsed since an arbitrary start time. The bottom segment is a plot of the magnetic flux
integrated over the area of the map as a function of time. The scale of $y$-axis in the bottom
segment is set at 10$^{18}$ Mx.}%
\label{elements}%
\end{figure*}

\section{SUMMARY AND DISCUSSION}\label{sec:disc}
With an aim to understand the role of the magnetic carpet in the heating of the solar corona, we
simulated the coronal magnetic field using the disk center observations of quiet Sun LOS magnetic
field obtained from the \textit{SDO}/HMI and \textit{Sunrise}/IMaX. To overcome the limitations of
short duration IMaX data, we created a time sequence of synthetic magnetograms that roughly match
the IMaX set in terms of the absolute flux (Figures~\ref{context}--\ref{timeseries}). A time
series of the 3D nonlinear force-free magnetic field is constructed with the magnetograms as lower
boundary conditions, and potential fields as initial conditions. The coronal field is evolved using 
Equation~\ref{eq:ind}, with a magneto-frictional relaxation technique (Equation~\ref{eq:mfr2}),
including hyperdiffusion (Equation~\ref{eq:hd}). We incorporated the non-force-free nature of the
lower solar atmosphere ($\leq0.7$ Mm) by adding vertical flows that prevent the field from splaying
out at the lower boundary (Equation~\ref{eq:mfr2}). These gentle flows weakly increased the field
strength of the magnetic elements.

Next, we calculated $E_{\text{free}}(z)$, $Q(z)$, and $F(z)$ to quantitatively estimate the
magnetic free energy, its dissipation, and energy flux injected at the base of corona. It is found
that the energy dissipation in simulations with moderate resolution (HMI), and much higher
resolution (IMaX/Source Model), is very similar close to the lower boundary. At low heights
in these models most of the heating occurs in short, low-lying loops that originate from the
mixed-polarity magnetic carpet. The average length of these loops is only a small fraction of the
computational domain. Further, these loops are highly intermittent, facilitating uniform energy
dissipation over the entire surface. These loops can be formed by interactions of near-by elements
\citep[e.g.][]{2010ApJ...723L.185W}, and also from the newly emerging flux in the
internetwork~\citep{2007ApJ...666L.137C}. 

Despite their intrinsic differences, IMaX and Source Model simulations are strikingly similar.
Moving away from the lower boundary, we see that $F_{\text{Source Model/IMaX}}$ decreases by more
than two orders of magnitude at $z=2$ Mm. This leaves only a few percent of $F$ at the coronal base.
The energy flux into the corona as predicted by our model is approximately $10^4$ $\rm erg ~
cm^{-2} ~ s^{-1}$ (see the right panel of Figure~\ref{eng_flux}), which is about one order of
magnitude smaller than the estimated radiative and conductive losses from the quiet-Sun corona as
derived from EUV and X-ray observations
\citep[e.g.,][]{1977ARA&A..15..363W,1988ApJ...325..442W,1991ApJ...382..667H}. We conclude that the
present model, which is based on evolving force-free magnetic fields and uses realistic photospheric
boundary conditions, cannot account for the observed coronal heating of the quiet Sun.

\begin{figure}
\begin{center}
\includegraphics[width=0.475\textwidth]{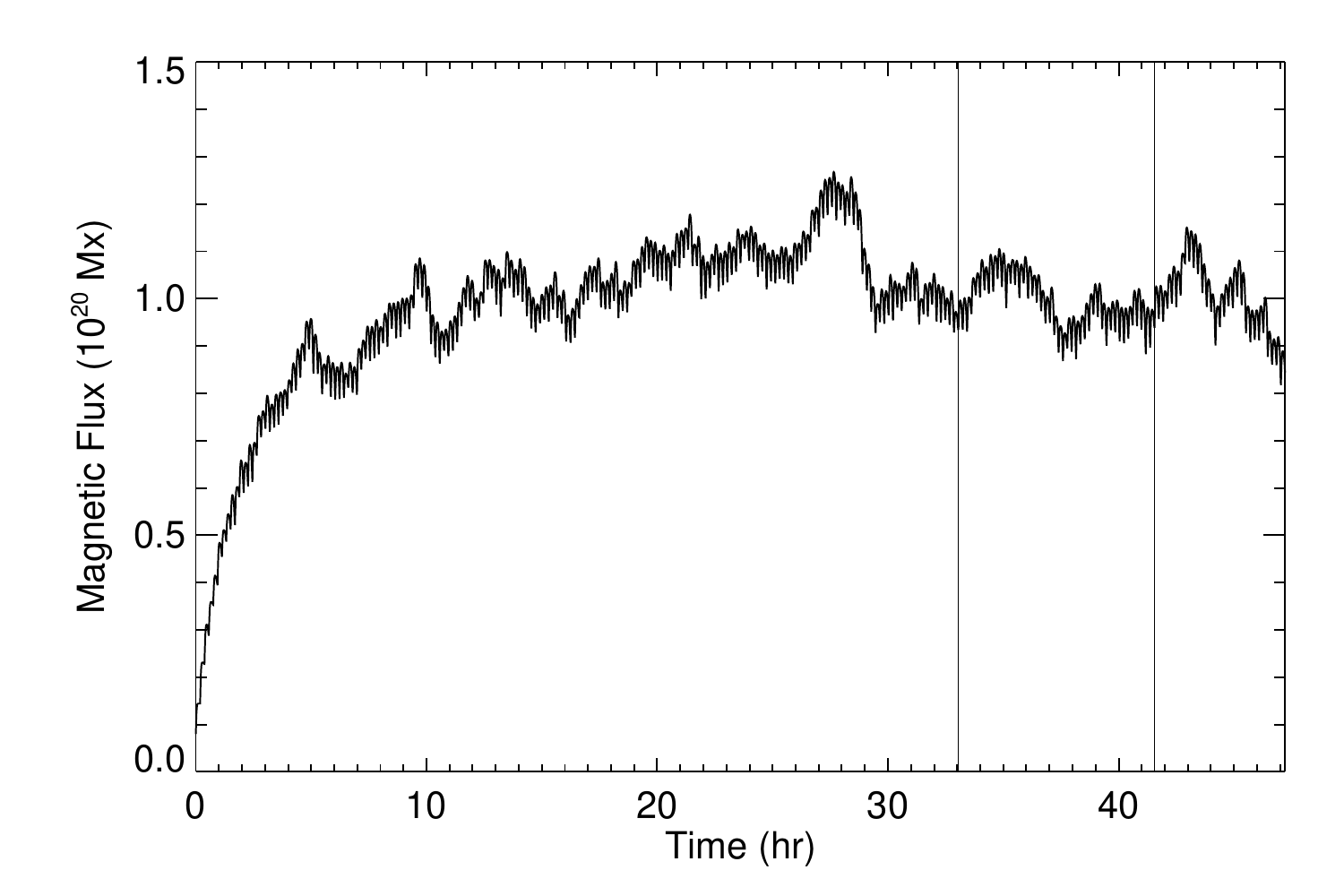}
\caption{The complete time evolution of the magnetic flux for the Source Model, integrated over its
FOV is shown. The plot includes both the rise and equilibrium phases of the magnetic flux as
described in Section~\ref{sec:dataset}. Two thin vertical lines mark the period of 8.5 hours
duration from which the time sequence of the Source Model is extracted for further
analysis.\label{source_flux}}
\end{center}
\end{figure}

\citet{2013SoPh..283..253W} used similar IMaX data to create a set of potential field models
of the evolving magnetic carpet. They also concluded that the magnetic energy release associated
with reconnections is not likely to supply the required energy to heat the chromosphere and corona.
Our estimate for the energy flux $F$ into the corona is below the upper limit provided by
\citet{2013SoPh..283..253W}, indicating that flux estimates based on force-free-field modeling are
consistent with those based on potential field modeling. However, both types of estimates are well
below the values needed to heat the corona. This suggests that both potential-field and
force-free-field models underestimate the amount of energy injected into the corona due to
photospheric footpoint motions and other effects in the magnetic carpet. This is perhaps not
surprising because these models make many simplifying assumptions about the structure and dynamics
of the solar atmosphere. For example, the models ignore the details of the lower atmosphere, where
the density decreases with height over many orders of magnitude. Also, the models assume that the
coronal magnetic field evolves quasi-statically in response to the footpoint motions, whereas in
reality the response may be more dynamic and significantly more energy may be injected in the form
of Alfv\'{e}n waves~\citep[e.g.,][]{2014ApJ...787...87V}.

However, the deficit in $F$ is smaller in the HMI set as compared to those of other models. Assuming
that this discrepancy is mainly due to the larger HMI volume\footnote{$6.6\times$ more grid cells
covering $192.4\times$ larger volume in physical space as compared to IMaX/Source Model volumes.},
it means that there is a contribution from the long range magnetic connections, often reaching
coronal heights, that was not possible with the IMaX set. This was also cited
by~\citet{2013SoPh..283..253W} as a reason for their too low upper limit. Future observations with
high spatial resolution and large FOVs are required to check and justify the contribution from long
range magnetic interactions. Also, IMaX and HMI observed mainly different kinds of features, with
very different lifetimes. In the case of HMI it is predominantly network magnetic features
(including ephemeral regions) and some internetwork fields, for IMaX it is dominated by internetwork
fields. The latter are more likely to be horizontal and do not reach high into the atmosphere.
Additionally, IMaX observations were carried out in 2009, at the depth of the last minimum in a very
quiet part of the Sun, while the HMI data were obtained in 2011, i.e. when the Sun was well on its
way into the new cycle.

A relatively large $F_{\text{HMI}}(z)$ can be considered as a \textit{basal} flux to heat a diffused
coronal region. Compact features like XBPs have a mean lifetime of eight hours
\citep{1974ApJ...189L..93G}, and an average emission height of 8\textendash 12 Mm above the
photosphere~\citep{2004A&A...414..707B,2007AdSpR..39.1853T}. Our model predicts a flux too
small to account for the energy losses at those height ranges.

Another important issue to scrutinize here is the lower boundary condition itself. Reliability of
the extrapolations using the photospheric magnetic field has to be further reviewed.
\citet{2007ApJ...665.1469A} demonstrated that the quiet Sun model chromosphere is a
non-force-free layer, and the extrapolations from the upper chromosphere may accurately yield the
quiet Sun field. Further, our simulations do not capture the actual response of plasma to the
dissipation, which is the determining factor when comparing the models with the emission maps of
corona. Also, it is possible that the energy released during events such as magnetic
reconnection at some altitude can be dissipated at a different altitude due to reconnection-driven
processes viz. jets, outflows, and waves~\citep{2004ESASP.575..198L}. These processes are not
included in the present simulations. Thus a detailed MHD description is necessary to evaluate the
role of the magnetic carpet in the heating of solar atmosphere. To conclude, our results show that
the energy flux associated with quasi-static processes (here modeled using magneto-friction and
hyperdiffusion) are not sufficient to heat the quiet solar corona. Although this suggests that the
evolution of the magnetic carpet (through magnetic splitting, merging, cancellation, and emergence)
without invoking any other reconnection-driven and wave mechanisms, does not play a
dominant role in the coronal energy supply, a final judgment can be passed only with better
observational constraints of magnetic field, in particular, at the force-free upper chromospheric
layer. Observations from the next generation instruments such as the Advanced Technology Solar
Telescope, and the Solar-C, which can offer multi-wavelength coverage with excellent
spectro-polarimetric sensitivity, and large FOVs, will shed light on these issues.

\acknowledgments 
The authors thank the Referee for many comments and suggestions that helped in improving the
presentation of the work. L.P.C. was a 2011\textendash 2013 SAO Pre-Doctoral Fellow at the
Harvard-Smithsonian Center for Astrophysics. Funding for L.P.C. and E.E.D. is provided by NASA
contract NNM07AB07C. Courtesy of NASA/SDO and the AIA and HMI science teams. This research has made
use of NASA's Astrophysics Data System.

\appendix
\section{Details of the Source Model}\label{source_model}
In this section we consider a model for the evolution of the photospheric magnetic field of the
quiet Sun. The field is assumed to consists of a collection of discrete flux elements or
``sources''. Each source has an associated magnetic flux, which can be positive or negative, but the
combined flux of all sources is assumed to vanish. Each source has a Gaussian flux distribution with
radius $r_0 = 200$ km. The spatial distribution of the sources continually evolves due to several
processes: (1) random motions driven by sub-surface convective flows, (2) splitting and merging of
like-polarity sources, (3) mutual cancellation of opposite-polarity sources, and (4) emergence of
new bipoles. To simulate these processes, we introduce a hexagonal grid with an edge length
$L\approx 1$ Mm. Initially, the sources are centered at randomly selected vertices of the hexagonal
grid (where 3 edges come together); subsequently they start moving along the edges of the grid. Each
source has an equal probability (0.25) of moving along one of the three edges connecting to its
original vertex, or remaining fixed at that vertex. The motion occurs with constant speed, $v = 1.5$
$\rm km ~ s^{-1}$, so that at time $t = \Delta t = L/v$ all moving sources again reach a neighboring
vertex. Then the process repeats itself, producing a random walk of the sources along the grid
edges. Therefore, in the present model all sources periodically return to the vertices (at times
$t_n = n\Delta t$ with $n = 1, 2, \cdots$).

The magnetic sources interact with each other only at vertices of the hexagons. For example, when
two sources meet at a vertex, they are forced to merge or partially cancel each other, depending on
their magnetic polarity. Also, each source has a probability of splitting in two, which only occurs
at a vertex so that the two parts may move away from each other along different edges. The emergence
of a bipole is modeled by inserting a new pair of flux-balanced sources at a vertex and allowing the
two sources to move apart along different edges of the grid. The newly inserted sources have
absolute fluxes in the range $0.1\text{\textendash}1.5 \times 10^{18}$ Mx.

Figure~\ref{elements} illustrates various interactions between sources: (a) flux emergence, (b)
emergence and subsequent partial cancellation, (c) splitting of an element, (d) total flux
cancellation. Each panel has two segments. The top segment shows the time sequence of the respective
interaction (the numbers given in the snapshots indicate time, in minutes, elapsed since an
arbitrary start time), and the bottom segment shows the magnetic flux vs.~time. Flux merging, and
self-cancellation of a newly emerged bipole are other possible ways in which the elements interact.
We emphasize that the motivation behind the Source Model is not to create a realistic magnetic flux
distribution for the quiet Sun, but rather to have a model with an average flux density similar to
that obtained in the IMaX observations.

The simulation is initiated with 50 magnetic sources with magnetic fluxes in the range
$5\text{\textendash}50 \times 10^{16}$ Mx. Figure~\ref{source_flux} shows the integrated flux of the
sources (i.e., the sum of absolute values of fluxes) as function of time. The initial phase of the
model is dominated by flux emergence. After 15 hours into the evolution, the integrated flux reached
a statistical equilibrium of $10^{20}$ Mx. This is due to the balance between the new flux emergence
and partial/complete cancellation of the magnetic elements. The two thin vertical lines mark the
period of evolution (8.5 hours) chosen for the main Source Model analysis presented in this work. In
total, 2000 bipoles emerged in the complete time evolution of the model (360 during the 8.5 hour
period). The average rate of flux emergence is about 150 $\rm Mx ~ cm^{-2} ~ day^{-1}$, which is a
factor three less than the value reported by~\citet{2011SoPh..269...13T}. However, their study
included a much wider range of flux emergence events ($10^{16}$\textendash$10^{23}$Mx). 


\end{document}